\documentclass[twocolumn,showpacs,superscriptaddress]{revtex4}
\usepackage{graphicx}
\usepackage{amsmath}
\usepackage{amsfonts}
\usepackage{amssymb}
\usepackage{color}
\usepackage{amscd}
\usepackage{bm}

\begin{document}
\title{Microscopic derivation of the Bekenstein-Hawking entropy for Schwarzschild black holes}
\author{Yong Xiao}
\email{yx234@sussex.ac.uk}
\affiliation{Key Laboratory of High-precision Computation and Application of Quantum Field Theory of Hebei Province,
College of Physical Science and Technology, Hebei University, Baoding 071002, China}
\affiliation{Department of Physics and Astronomy, University of Sussex, Brighton, BN1 9QH, United Kingdom}

\begin{abstract}
In this paper, we successfully derive the Bekenstein-Hawking entropy for Schwarzschild black holes in various dimensions by using a non-trivial phase space. It is appealing to notice that the thermodynamics of a Schwarzschild black hole actually behaves like that of a $1$-dimensional quantum mechanical system.
Our result suggests that black hole should be viewed as a system with the equation of state $P=\rho$, and it also suggests that a holographic
stage should exist in the early universe.
\end{abstract}
\pacs{04.70.Dy, 11.10.Cd, 03.70.+k} \maketitle

\section{Introduction}
Though black holes are originated from classical solutions to the Einstein field equation, it
is well established that they have thermodynamical behaviors such as temperature and entropy.
The famous Bekenstein-Hawking entropy takes the form
\begin{align}
S_{BH} = k_B \frac{A}{4 l_p^2}, \label{bh}
\end{align}
where $l_p=\sqrt{ \hbar G /c^3 }$ is the Planck length. This form of entropy is often called holographic entropy,
for that it is proportional to the boundary area of the system.
The microscopic origin of the holographic entropy has always been a question to be answered.
The presence of $K_B$, $\hbar$, $c$ and $G$ in eq.\eqref{bh} implies that its explanation should
involve statistical mechanics, quantum mechanics, special relativity and gravitational physics.

It has been known that conventional quantum field theory (QFT) cannot provide enough degrees of freedom
to account for the holographic entropy. The entropy bound for conventional QFT under gravitational constraint is
$ k_B ( \frac{A}{ l_p^2} )^{\frac{3}{4}}$ \cite{hooft,bousso,cohen,hsuEB,us1,hsuMN,barrow}. Obviously there is
a huge entropy gap between the maximum entropy of conventional QFT and the
holographic entropy of black holes. An immediate question is that what kind of microscopic theory can account for
 the holographic entropy? And in what aspects should the theory be distinct from the conventional QFT?

Note that what we have stressed is that black hole physics cannot be described by a conventional \textit{bulk} QFT,
it does not conflict with the idea of AdS/CFT which is a correspondence between theories in different space-time dimensions.
Though AdS/CFT has gained many achievements by attaching the properties of certain black holes with CFTs in lower dimensions,
 it is still worthy to gain more understandings about the bulk theory itself and to explain the microscopic structure of black hole directly.
In addition, there are surely many problems to be solved which cannot fit into the framework of AdS/CFT easily, such as the
entropy of the Schwarzschild black hole \cite{bekenstein}, which is far from being extremal and lives in an asymptotic flat space-time,
and the cosmological entropy bounds \cite{barrow,fs}. Our work may provide some new insights into these problems.

The paper is organized as follows. We first review the derivation of the maximum entropy of conventional QFT as a preparation. Then we manage
to derive the area-scaling entropy for quantum gravitational systems by simple dimensional analysis and gain insights about of the microscopic
physical laws behind it. Based on a non-trivial phase space structure, we derive the exact Bekenstein-Hawking entropy for Schwarzschild black holes
in various dimensions and discuss the corresponding microscopic pictures. Finally, we make a discussion about the implication of our result
to black hole physics and cosmology.

\section{The maximum entropy of conventional quantum field theory}
The entropy bound $ k_B ( \frac{A}{ l_p^2} )^{\frac{3}{4}}$ for conventional QFT under gravitational constraint was first derived by 't Hooft in \cite{hooft}, and has been
 verify by various approaches \cite{bousso,cohen,hsuEB,us1,hsuMN,barrow}. We review the derivations to the entropy bound and
show that dimensional analysis is enough to get the correct scaling behavior of the entropy bound,
while concrete microscopic physics can provide exact coefficients to the relevant formulas.

Consider a typical QFT system of size $L$, and take the average energy of each particle inside the system to be $k_B T$.
Then, by simple dimensional analysis, the energy and entropy of the system can only take the form
\begin{align}
E\sim L^3 T^4, \  \   \ S \sim L^3 T^3.  \label{eeesss}
\end{align}
We did not introduce the mass parameter $m$ into the expressions, because
a system consisting of massless particles always has
more entropy than their massive partners with the same relativistic energy.

Imposing the gravitational constraint that the energy of the system does
not exceed the energy of a black hole of the same size,
$E\sim L^3 T^4\leq E_{BH} \sim L $, one easily gets the maximum realizable temperature $T_{max}
\sim L^{-1/2}$. Substituting it into the entropy formula, the maximum entropy is
\begin{align}
S_{max} \sim L^{ \frac{3}{2} } \sim A^{ \frac{3}{4} },\label{bdd}
\end{align}
where $A$ is the boundary area of the system.

Actually, due to our knowledge of conventional QFT, it is easy to provide a microscopic
derivation of this entropy bound \eqref{bdd}. When bosonic quantum fields are confined inside a
box, the basic modes of the system can be listed as $\vec{p}_i=\frac{2\pi \hbar}{L} (m_x,m_y,m_z)$, where $m_x$, $m_y$, $m_z$
are quantum numbers labeling the mode. Acting the corresponding creator operators $a_{p_{i}}^{\dag}$ on the vacuum state $|0\rangle$,
the quantum states of the system can be listed as
\begin{align}
|\psi_s\rangle= \cdots   \left(
a_{p_{i}}^{\dag}\right)^{n_{i}}   \cdots      \left(
a_{p_{2}}^{\dag }\right) ^{n_{2}}\left(
a_{p_{1}}^{\dag}\right)^{n_{1}}|0\rangle.\label{state}
\end{align}
In a field-theoretical language, $n_i$ particle are excited on the $i$-th mode, and
different sets of the occupation number $\{n_i\}$ corresponds to different microscopic states of the system. Assume the gravitational constraint
\begin{align}
E_{ |\psi_s\rangle }= \sum\limits_{i}    n_{i}\varepsilon_{i}\leq
E_{bh}, \label{4}
\end{align}
where $\varepsilon_i=c p_i= \frac{2\pi \hbar c}{L}\sqrt{m_x^2 + m_y^2+m_z^2}$ is the energy attached to each mode. Then the total number of the quantum states satisfying this limitation \eqref{4} can be counted and proven to be
$W\sim e^{ (A/ l_p^2 )^{3/4} }$ \cite{us1}. The direct counting method has the advantage that independent quantum states are listed clearly and it corresponds
 to the micro-canonical ensemble method in statistical mechanics.

In most cases, canonical ensemble method is more convenient by boiling the question down to the calculation of partition function. Taking photon gas system for example, the logarithm of the partition function is given by \cite{huang}
\begin{align}
\begin{split}
\ln \Xi &= - \sum\limits_{i} \ln ( 1 - e^{ - \beta \varepsilon_i} )
  \\ &=  \frac{2V}{ (2\pi \hbar)^3 }  \int  \ln ( 1 - e^{ -
\beta cp }  ) d^3\vec{p}
=\frac{ \pi^2}{45 c^3 \hbar^3 } \frac{V}{\beta^3},
 \end{split} \label{lnxiqft}
\end{align}
where $\beta=1/k_B T$ and the summation over independent modes is evaluated by the volume of phase space $V d^3\vec{p}$  divided by $(2\pi \hbar)^3$.
It follows the energy and entropy of the system as
\begin{align}
 E= - \frac{\partial }{\partial \beta}  \ln \Xi =\frac{\pi^2 k_B^4 }{15 c^3 \hbar^3 }V T^4, \label{QFTe} \\
S= k_B (\ln \Xi  + \beta E ) =\frac{4 \pi^2 k_B^4}{45 c^3 \hbar^3} V T^3, \label{QFTs}
\end{align}
along with the equation of state $P=\frac{1}{3}\rho$. Comparing them to eq.\eqref{eeesss} derived from dimensional analysis, the microscopic physics of photons only determines the exact coefficients. Imposing $E\leq E_{bh}$, the exact entropy bound for photon gas can be readily obtained. One can also refer to \cite{sorkin,chav} for a detailed analysis of the self-gravitating photon system and the corresponding behavior $E\sim L$, $T\sim L^{-1/2}$ and $S\sim A^{3/4}$. It is conceivable that, when gravity is extremely strong and takes over the system, the basic relations \eqref{eeesss} must be greatly modified. Then we say the conventional QFT is no longer applicable and a new type of theory is needed.

\section{Area-scaling entropy by dimensional analysis}
Black hole thermodynamics is expected to be explained by a microscopic quantum gravitational theory. But at present
we do not know enough about the fundamental principles of such a theory. Fortunately, as have been noticed,
simple dimensional analysis is rather useful in determining the scaling behaviors of the thermodynamical quantities and reveals
some information of the microscopic physics.

Since the gravitational Hamiltonian derived from Einstein-Hilbert action is proportional to $1/G$ and the volume integral, it is natural
to conjecture the energy and entropy of a quantum gravitational system as
 \begin{align}
E\sim  \frac{1}{G} V T^2, \ \ \ \   S \sim  \frac{1}{G} V T, \label{dim}
\end{align}
where $V$ and $T$ are respectively the volume and temperature of the system. Moreover, in the spirit of dimensional analysis,
 we do not need to worry about the effect of might-be highly curved space-time, unless the space-time
is so curved to produce a new characteristic length scale. Now requiring the energy to be $E_{bh}\sim L$, it follows immediately $T\sim L^{-1}$ and $S\sim A$.
 So it is easy to derive the scaling behaviors of black hole thermodynamics.

Assume the system consists of some microscopic particles, which may be gravitons or some unknown particles but would not be photons again.
We want to know whether the formulas \eqref{dim} from dimensional analysis could reveal some microscopic physical principles to us.
Go back to the conventional QFT case to get some inspirations. Obviously, in the formula \eqref{lnxiqft}, the speed of light $c$ plays the role of attaching energy and momentum of the photons, that is, $\varepsilon=c p $.
And the Planck constant $\hbar$ comes from the quantum uncertainty principle $\triangle q_i \triangle p_i \geq \frac{\hbar}{2}$ which is the basis
of using $\frac{V dp^3}{ (2\pi \hbar)^3 }$ to count the independent quantized modes. Now turn to analyze the quantum gravitational case. The complete form of the entropy in eq.\eqref{dim} can be suggestively written as
\begin{align}
 S \sim \frac{ c^2 }{G \hbar^2}VT \sim \frac{1}{ \hbar c } L_{s} T, \label{sss}
 \end{align}
  with $L_{s}\equiv\frac{V}{l_p^2}$. Compare eq.\eqref{sss} with $S\sim \frac{1}{\hbar^3 c^3} V T^3$ of the conventional QFT case carefully. It appears to us that we were studying a $1$-dimensional quantum system other than a $3$-dimensional quantum system in some sense. Concretely speaking, we can still use $c$ to attach energy and momentum. But, in order to mathematically derive the correct form of eq.\eqref{sss}, we must use $\frac{L_s dp}{2\pi  \hbar} $ to count the number of quantized modes in the quantum gravitational case, other than $\frac{V d^3 \vec{p} }{ (2\pi \hbar)^3 }$ in the conventional quantum QFT case.

The non-trivial quantum phase space structure surely implies a drastic modification of quantum uncertainty relation and the basic quantum commutation relation.
But we temporarily concentrate on our statistical derivation of holographic entropy and go back to this issue later on.

\section{A microscopic derivation to the Bekenstein-Hawking entropy}
Based on our analysis above, we make the following assumption of the microscopic particles inside a quantum gravitational system. First, the particles are massless and bosonic. Second, they obey the energy-momentum relation $\varepsilon=c p$ with $p=|\vec{p}|$. Third, the number of independent quantized modes should be evaluated by $g \frac{ L_s dp}{2\pi  \hbar}=g\frac{ c^3 V dp}{ 2\pi G \hbar^2 }$, other than the conventional $\frac{V d^3\vec{p}}{ (2\pi \hbar)^3 }$. Here a dimensionless coefficient $g$ is introduced to include other possible degrees of freedom such as polarization.

We model the Schwarzschild black hole of radius $R$ in $3+1$ dimensions as a system consisting of these particles. All those calculations for photon system can be parallel translated to the new system, except for the non-trivial quantum phase space. Though the scaling behaviors of the thermodynamical quantities have been reserved in advance, it is hard to believe one can get the exact Bekenstein-Hawking entropy from such a simple setting before doing calculation.

Now the logarithm of the partition function is
\begin{align}
\ln \Xi =-\frac{g  c^3 V}{\pi G \hbar^2} \int_0^\infty  \ln \left( 1 - e^{ -
\beta cp }  \right) dp= \frac{g \pi c^2}{6 G \hbar^2} \frac{V}{\beta}, \label{lnxho}
\end{align}
where $V=\frac{4\pi R^3}{3} $ \footnote{Here is a subtlety. The phase space volume $d^3\vec{x} d^3\vec{p}$ is actually invariant under a general transformation of coordinates. (A direct analysis of the Jacobian of the transformation can be found on page $129$ of the lecture ``General Relativity and Cosmology" by Guido Cognola.) Thus there is no need to introduce an extra term like $\sqrt{\gamma}$ to include the influence of the possible curved space, as one may worry. Instead, we should assume a non-trivial energy spectrum to make difference with the conventional QFT case. By hindsight, in eq.\eqref{eqvv} where $p_e$ has a usual explanation, the non-trivial energy spectrum is $\varepsilon=\frac{ G}{c^2 \hbar} p_e^3 $, and it can transform to the non-trivial phase space in eq.\eqref{lnxho} by a change of variable. }.
 Then we get the expressions for the energy and entropy as
\begin{align}
E = - \frac{\partial }{\partial \beta}  \ln \Xi = \frac{g \pi k_B^2 c^2}{6 G \hbar^2} V T^2,  \label{energy} \\
S =  k_B (\ln \Xi  + \beta E) =  \frac{g \pi  k_B^2 c^2}{3 G \hbar^2}  V T. \label{entropy}
 \end{align}
The pressure of the system can be calculated as
\begin{align}
P =
k_B T \frac{\partial \ln \Xi }{\partial V} =\frac{g \pi k_B^2 c^2}{6 G \hbar^2}  T^2.
\end{align}
Comparing with $\rho=E/V$, we find the equation of state of the system as
\begin{align}
P=\rho.
\end{align}
The Komar mass as the gravitational source corresponds to $(\rho+3P)V$,
so we get
\begin{align}
M=4E=\frac{2g \pi k_B^2 c^2}{3 G \hbar^2} V T^2.\label{komar}
\end{align}
 Taking $M$ to be the energy of the black hole $M=\frac{c^4}{2G}R$ and choosing $g=9$ in eq.\eqref{komar}, we deduce the exact Hawing temperature $T=\frac{\hbar c}{4\pi k_B}\frac{1}{R}$ \footnote{Conversely speaking, the mass and the Hawking temperature together can fix $g=9$ and lead to the exact Bekenstein-Hawking entropy. The emergence of the number $g=9$ is intriguing. This means, if we view the system as a quantum mechanical string as suggested in next section, it actually has $9$ directions to vibrate. Thus the $3+1$ dimensional Schwarzschild black hole can be viewed as a string living in $9+1$ dimensions (or $10+1$ dimensions if the string can only vibrate along transverse dimensions). The extra $6$ dimensions may be curled up to fit with the $3+1$ dimensional phenomena.}. Substituting this into eq.\eqref{entropy}, we finally get the exact Bekentstein--Hawking entropy
 \begin{align}
 S= k_B \frac{A}{4l_p^2},
 \end{align}
 which surely has a statistical interpretation. Note that another reason of using $M$ other than $E$ is that $dE+PdV=TdS$ can be written as $dM=TdS$ in order to fit with the thermodynamics of the Schwarzschild black hole \footnote{There is $E\sim R$ for a general self-gravitating system, so only one independent variable is needed to describe this kind of systems. It is easy to check $dE+PdV=d (E+3PV)$ with $P=w\rho \sim R^{-2}$ and $V\sim R^3$. For a $D=d+1$ dimensional self-gravitating system, the result is $dE+PdV=d (E+\frac{D-1}{D-3} PV)$. Accordingly, we can identify $M=E+\frac{D-1}{D-3} PV$ in the formula $dM=TdS$.}.

In fact, by comparing eqs.\eqref{entropy} and \eqref{komar}, we can go straightly to observe the relation
\begin{align}
TS=\frac{1}{2}M,\label{smarr}
\end{align}
which is exactly the same as the Smarr formula for $3+1$ dimensional Schwarzschild black hole. Reading $M=\frac{c^4}{2G}R$ and $T=\frac{\hbar c}{4\pi k_B}\frac{1}{R}$ directly from the Schwarzschild metric and substituting them into eq.\eqref{smarr}, the exact Bekentstein--Hawking entropy follows as
 \begin{align}
S= \frac{M}{2T} = k_B \frac{\pi R^2}{l_p^2} = k_B\frac{A}{4l_p^2}.
\end{align}

To remove possible doubts, we further generalize the above derivation to higher dimensional Schwarzschild black holes.
In $D=d+1$ dimensional space-time, the partition function is
\begin{align}
\begin{split}
\ln \Xi &=-\frac{g_D  c^3 V_{D-1} }{\pi G_D \hbar^2} \int_0^\infty  \ln \left( 1 - e^{ -
\beta cp }  \right) dp \\&= \frac{g_D \pi  c^2}{6 G_D \hbar^2}  \frac{V_{D-1}}{\beta}.
\end{split}
\end{align}
It takes the same form as eq.\eqref{lnxho} with only $g$, $G$ and $V$ changed to their higher-dimensional counterparts $g_D$, $G_D$ and $V_{D-1}$. Thus the same formulas are derived as
\begin{align}
TS=2E,\ \ \ P=\rho. \label{enss}
\end{align}
The gravitational source $M$ in $D$-dimensional space-time corresponds to $(\rho + \frac{D-1}{D-3} P)V_{D-1}$.
Because of $P=\rho$, there is $M=\frac{2(D-2)}{D-3}E$. Comparing with eq.\eqref{enss}, we get the relation
\begin{align}
TS=\frac{D-3}{D-2}M,
\end{align}
which is the same as the Smarr formula for Schwarzschild black holes in general dimensions.
Needless to say, substituting the mass and Hawking temperature of the black hole into it,
we get the exact Bekenstein--Hawing entropy
\begin{align}
S= k_B\frac{A_{D-2}}{4 l_p^{D-2}},
 \end{align}
 which is more than could be expected. In the above derivation, one may have noticed that the equation of state $w\equiv\frac{P}{\rho}=1$ is critical to derive the Smarr formulas. If using other value of $w$, one will end up with a wrong coefficient. Take $w=\frac{1}{3}$ in $3+1$ dimensional space-time for example, the best result that one can get is $TS=\frac{2}{3}M$ and $S=\frac{4}{3}S_{BH}$ \footnote{To circumvent the $A^{3/4}$ entropy bound, one should modify eqs.\eqref{QFTe} and \eqref{QFTs} to $E\sim zL^3 T^4$, $S\sim zL^3 T^3$, where $z$ represents the number of different types of massless particles with $w=1/3$ in conventional QFT and it virtually serves as a new source of entropy. Choose $z\sim L^2/l_p^2$, which is so large to change the behaviors of the system to $E\sim L^5 T^4$, $S\sim L^5 T^3$. Then one can realize the expected scaling behavior $M\sim L$, $T\sim L^{-1}$ and $S\sim A$ but an extra factor $4/3$ to the Bekenstein--Hawking entropy. Note there is a similar $4/3$ problem in the context of AdS/CFT which may be cured by considering strong coupling of the fields \cite{gubser}.}. Clearly, only the microscopic physics with $w=1$ leads to the exact Bekenstein--Hawking entropy.

 \section{More on the microscopic picture}
 Though we have successfully derived the Bekenstein--Hawking entropy, we did not say too much about the concrete microscopic picture of the system. Actually, the concrete microscopic picture of the system strongly depends on how to interpret the nontrivial phase space structure.
Below we shall conjecture two possible microscopic pictures, which are equivalent to each other in the sense that they lead to the same partition function.

The first picture is that the quantum gravitational system with volume $V$ can be viewed as a $1$-dimensional quantum mechanical system with
length  $L_{s}\equiv\frac{V}{l_p^2}$, as suggested from the form of eq.\eqref{sss}. The length $L_s$ is far longer than the size of the black hole, so it would be interesting
to imagine it as a very long non-relativistic string highly curling and winding inside the system. The energy of the corresponding modes is quantized as $\varepsilon_i=\frac{2\pi \hbar c} {  L_{s} }m_i $, with $m_i=1,2,3\cdots$.
Then all the quantum states of the system can be described by
\begin{align}
|\psi_s\rangle= \cdots   \left(
a_i^{\dag}\right)^{n_{i}}   \cdots      \left(
a_2^{\dag }\right) ^{n_{2}}\left(
a_1^{\dag}\right)^{n_{1}}|\Omega\rangle. \label{statess}
\end{align}
These $N=n_1+n_2+\cdots $ excitations on the string provide the fundamental particles inside the quantum-gravitational system.
The partition function of the system is exactly that given by eq.\eqref{lnxho}. In fact some important properties of the system can be easily
observed, for example, the equation of state of the system must be
\begin{align}
P =  - \frac{\partial}{\partial V} (\sum\limits_i {n_i \varepsilon
_i })  =\sum\limits_i { \frac{n_i \varepsilon _i}{V} } = \frac{E}{V}=
\rho,
\end{align}
by noting that $\frac{\partial \varepsilon_i} {\partial V}= -\frac{\varepsilon_i}{V}$ due to $\varepsilon\sim 1/L_s \sim 1/V$.
Furthermore, the number of quantum states \eqref{statess} satisfying $E =\sum\limits_{i}  n_{i}\varepsilon_{i}\leq
E_{bh}$ can be easily counted out by writing it in the form $\sum\limits_{i} n_{i} m_{i}\leq \frac{ E_{bh} L_s}{2\pi \hbar c}$.
In mathematics it refers to the integer partition problem, that is, counting the number of different ways of writing a large number as a sum of positive integers.
By using the Hardy-Ramanujan partition formula, we get the number of permitted quantum states of the system as $W\sim e^{\frac{A}{4l_p^2} }$.
Still, the fact $w=1$ is essential in the derivation to get the exact coefficient $\frac{1}{4}$. It is amazing to see that Bekenstein--Hawking entropy can be derived from such a simple picture.

In the second picture, we try to maintain the $3$-dimensional uncertainty relation. Now stare at the non-trivial quantum phase space $\frac{c^3 Vdp}{G\hbar^2}$.
Obviously, if we introduce some effective momentum $\vec{p}_{e}$ satisfying $p\equiv\frac{ G}{c^3 \hbar} p_e^3$, we will recover the
normal behavior of phase space $\frac{Vp_e^2 dp_e}{ \hbar^3 }$ or written clearly as  $\frac{ d^3 \vec{x} d^3 \vec{p}_e }{(2\pi \hbar)^3}$. Thus the effective momentum
$\vec{p}_e$ has the same quantum uncertainty relations as the normal momentum in conventional QFT and it should be quantized as usual $\vec{p}_e= \frac{2\pi \hbar}{L}(m_x,m_y,m_z)$. Then the quantum states of the system can also be listed as the form \eqref{state}. The only difference is that the modes are attached with a weird energy $\varepsilon=cp=\frac{ G}{c^2 \hbar}  p_e^3$. Accordingly, the logarithm of the partition function is
 \begin{align}
\ln \Xi \sim -\frac{ V  }{ (2\pi \hbar)^3 }  \int_0^\infty  \ln \left( 1 - e^{
- \beta \varepsilon }  \right) p_e^2 dp_e. \label{eqvv}
\end{align}
This is actually eq.\eqref{lnxho} with a change of variable. In this picture, for a black hole with average particle energy $\varepsilon\sim k_B T\sim \hbar/R$, we should use $p_e$ other than the obscure $p$ to
calculate the characteristic thermal wavelength $\lambda$ of the system, which gives $p_e \sim l_p^{-2/3} R^{-1/3}$ and $\lambda \sim l_p^ {2/3} R^{1/3}$. It means each independent wave-packet inside black holes occupies a volume $\lambda^3 \sim l_p^2 R$. The
interesting part is that, by comparing with the equations of van der Waals
fluids, the specific volume of
the ``molecules'' constituting charged AdS black boles is exactly
identified as $2 l_p ^2 R$ with $R$ the horizon size \cite{kub,wei,wei2}. Besides, the length uncertainty of measuring a distance $L$ has also been identified as a similar form $\delta L=l_p^{2/3} L^{1/3}$ based on quantum mechanical and gravitational principles \cite{fk,mazi}. It is not clear whether there are some deep connections here.

\section{Conclusion and Discussions}
In this paper, we have successfully provided a microscopic derivation of the Bekenstein--Hawking entropy for Schwarzschild black holes in various dimensions, by considering the black holes as consisting of microscopic particles with a nontrivial phase space. It is appealing to note that the thermodynamics of a Schwarzschild black hole resembles that of a quantum mechanical non-relativistic string. In fact, the microscopic physical contents inside some black holes have been conjectured and discussed from various motivations in recent years \cite{wei,wei2,polchinski,pol2,vene,maj}, in sharp contrast to a classical picture of black holes with empty interior expect for a singularity at the centre. For example, charged AdS black holes are suggested to be consisted of ``molecules'' with attractive or repulsive interactions \cite{wei,wei2}. The feature of our present work is that the Schwarzschild black hole may be viewed as a very long quantum mechanical string at least mathematically and the exact Bekenstein--Hawking entropy can be derived.

The idea that black holes might be $1+1$ dimensional in some sense can be traced back to \cite{bema,nation,majhi,carli}. In the seminal paper \cite{bema} Bekenstein and Mayo found that the relation between the entropy flow rate $\dot{S}$ and the power $P_r$ of radiating energy of a $3+1$ dimensional black hole is $\dot{S}\sim \sqrt{P_r}$, the same as that of a one dimensional information channel. Thus they judged that a black hole is effectively $1+1$ dimensional as far as entropy flow is concerned. Actually we can easily generalize their derivation to $D=d+1$ dimensional black holes, where $\dot{S}\sim A_{D-2} T^{D-1}\sim 1/R$ and $P_r\sim A_{D-2} T^{D}\sim 1/R^2$. As a result there is always $\dot{S}\sim \sqrt{P_r}$. This nicely favors our work viewing black holes in general dimensions as a $1+1$ dimensional system.

Our work also suggests that the black hole can be viewed as a massive object with equation of state $P=\rho$. It would be interesting to consider this equation of state in the study of the phenomena of black hole coalescence and to see whether or not it would make differences in the numerical simulations and gravitational-wave observations. After finishing this work, we notice that there have been a lot of interesting researches based on the fluid with $P=\rho$. So we make some discussions about our work and the existing literature below.

Interestingly, the connection between the equation of state $P=\rho$ and the black hole entropy has been noticed decades ago \cite{zp,tho}.  The authors
managed to solve the Tolman-Oppenheimer-Volkoff (TOV) equation with $P=w\rho$, and the entropy was calculated as the integral of $\beta (\rho(r) + P(r))$ while taking $\beta$ to be the inverse of Hawking temperature. Though there must be a negative-mass singularity at the centre of the star after fitting with an exterior Schwarzschild metric and the interior metric is abnormal in some regions, they found the entropy becomes $S=k_B \frac{A}{4l_p^2}$ when taking $w=1$. However, the fundamental reason why the entropy could emerge from such tedious calculations were not clearly understood. By comparison,  $P=\rho$ is a derived result from our microscopic picture. In our opinion, the key to understanding their success is that $w=1$ has implicitly equalized their entropy $S$ and $\frac{1}{2}\beta M$ (that is, $1+w$= $\frac{1}{2}(1+3w)$ when $w=1$). If their expressions can give the required $M$ as the black hole, which more or less is constrained by the boundary condition of the TOV equations, the Bekenstein--Hawking entropy follows. In addition, the presence of the negative-mass singularity in their scenario may imply that the classical description along with the TOV equation is not enough to deal with the relevant physical situation. Quantum mechanics may need to be fully considered to exclude the singularity. It is also possible that a self-consistent description may have to include extra dimensions or wormholes \cite{polchinski,malsuss} in order to avoid the singularity. These possibilities are worthy of further study.

In the context of cosmology, we mainly concern about which stage the holographic fluid with $w=1$ may dominate in the history of the universe.
 First, by Friedmann equations the evolvement of the universe declines to lower the value of $w$ as time increases. So it is natural to expect an early stage of the universe with $w=1$ before the radiation dominated universe with $w=\frac{1}{3}$. Second, when tracing back the history of the universe, we encounter from atomic physics to nuclear physics
and to grand unified physics. Our work suggests $w=1$ is closely related to quantum gravity and holographic entropy,
so it provides another independent logic to the same conclusion that a $w=1$ stage should exist before the conventional QFT dominated
stage of the universe. Actually, we find that the fluid with $w=1$ has already been conjectured and studied in cosmology for many years \cite{zel}.
It is usually called stiff fluid in the literature, for that it is the most incompressible fluid permitted by relativistic causality.
Such a kind of fluid surely has a large number of possible physical origins \cite{dut,chav2,nai} different from what we have suggested.
Interestingly, there are also a series of works called ``holographic cosmology" \cite{bf1,bf2}, since after Fischler and Susskind
showed the cosmologic holographic entropy bound could be saturated by $w=1$ \cite{fs}. Even a holographic eternal inflation model
has been put forward \cite{bf4}. Thus, if we take seriously about the holographic stage with $w=1$, the understanding of the early universe
including the picture of the Big Bang and inflation might be greatly modified. We hope the remnant indications of this holographic stage could be detected in
future cosmological experiments \footnote{Note added. After the paper was posted on arXiv, I was informed about the relevant works \cite{b1,b2,b3} where it is suggested that the interior of black hole must be in some kind of highly quantum state and the Schwarzschild black hole is viewed as a bound state of highly exited, long, closed strings with interactions between them. In contrast, our scenario here is a single long string with length $L_s\sim V/l_p^2$ and massless particles excited on it. I also noticed a large number of earlier works which attach $\rho=p$ with black hole thermodynamics, for example, see \cite{mathur} and references therein. It would be interesting to compare the similarity and differences to find deep insights about understanding the microscopic structure inside black hole.}.

\section*{Acknowledgements} I would like to thank X. Calmet for useful discussions. I am grateful to MPS School of the University of Sussex for the research facilities and their hospitality during my visit. The work is supported by China Scholarship Council (No. 201908130079) and the Optical Engineering Key Subject Construction Project of Hebei University.

\end{document}